\documentclass[fleqn,twoside]{article}
\usepackage{espcrc2}
\input psfig
\title{Solar Models: An Historical Overview}

\author{John N. Bahcall\address{School of Natural Sciences,
Institute for Advanced Study, Princeton, NJ,
USA}\thanks{jnb@ias.edu}}

\begin{document}

\begin{abstract}
I will summarize in four slides the $40$ years of development of
the standard solar model that is used to predict solar neutrino
fluxes and then describe the current uncertainties in the
predictions. I will dispel the misconception that the p-p neutrino
flux is determined by the solar luminosity and present a related
formula that gives, in terms of the p-p and $^7$Be neutrino
fluxes, the ratio of the rates of the two primary ways of
terminating the p-p fusion chain. I will also attempt to explain
why it took so long, about three and a half decades, to reach a
consensus view that new physics is being learned from solar
neutrino experiments. Finally, I close with a personal confession.
\vspace{1pc}
\end{abstract}

\maketitle
\section{Introduction}
\label{sec:introduction}

I will follow in this text the content of my talk at Neutrino2002,
which occurred  in  Munich, May 25-30, 2002.

I begin in Section~\ref{sec:ray}, as I did in Munich, with a
tribute to Ray Davis. In Section~\ref{sec:development}, I present
a concise history of the development of the standard solar model
that is used today to predict solar neutrino fluxes. This section
is based upon four slides that I used to summarize the development
and is broken up into four subsections, each one of which
describes what was written on one of the four slides. I describe
in Section~\ref{sec:uncertainties} the currently-estimated
uncertainties in the solar neutrino predictions\footnote{Where
contemporary numbers are required in this review, I use the
results from the BP00 solar model, ApJ 555 (2001) 990,
astro-ph/0010346.}, a critical issue for existing and future solar
neutrino experiments. I show in Section~\ref{sec:ppflux} that the
solar luminosity does not determine the p-p flux, although there
are many claims in the literature that it does. I also present a
formula that gives the ratio of the rates of the $^3$He-$^3$He and
the $^3$He-$^4$He reactions as a function of the p-p and $^7$Be
neutrino fluxes. These reactions are the principal terminating
fusion reactions of the p-p chain. In Section~\ref{sec:why},I give
my explanation of why it took so long for physicists to reach a
consensus that new particle physics was being learned from solar
neutrino experiments. I close with a personal confession in
Section~\ref{sec:confession}.

\section{Ray Davis}
\label{sec:ray}

\begin{figure}[!ht]
\centerline{\psfig{figure=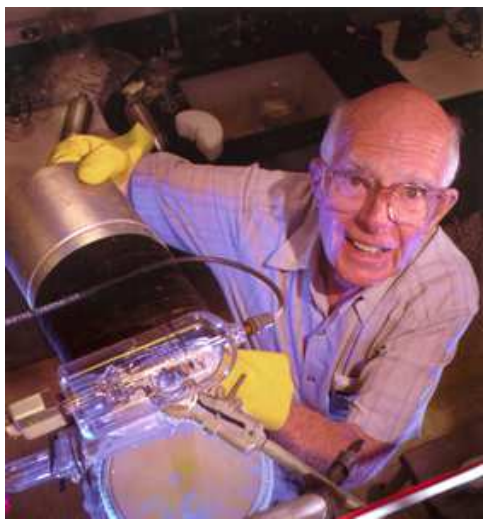,width=2.5in}} \vglue-.3truein
\caption[]{Ray Davis preparing to pour liquid nitrogen into a
dewar on a vacuum system of the type used for gas purification and
counter filling in the chlorine experiment. The glass object in
the foreground with the wire coming out that blocks Ray's left
hand is an ionization gauge used to measure the pressure in the
vacuum system.\label{fig:ray}}
\end{figure}

Before I begin the discussion of the standard solar model, I would
like to say a few words about Ray Davis, shown in
Figure~\ref{fig:ray}. The solar neutrino saga has been a community
effort in which thousands of chemists, physicists, astronomers,
and engineers have contributed in crucial ways to refining the
nuclear physics, the astrophysics, and the detectors so that the
subject could become a precision test of stellar evolution and,
ultimately, of weak interaction theory.

However, Ray's role in the subject has been unique. Any historical
summary, even of solar models, would be grossly incomplete if it
did not emphasize the inspiration provided by Ray's experimental
vision. Although Ray never was involved in solar model
calculations, and has always maintained a healthy skepticism
regarding their validity, his interest in performing a solar
neutrino experiment was  certainly the motivation for my entering
and remaining in the subject. More importantly, for all of the
formative years of the ``solar neutrino problem", Ray inspired
everyone  who became involved with solar neutrinos by his
conviction that valid and fundamental measurements could be made
using solar neutrinos. We committed to a subject that did not
attract main stream scientists because we believed in Ray's dream
of measuring the solar neutrino flux.

Ray and I have written three articles on the history of solar
neutrino research (in 1976, 1982, and 2000, see
http://www.sns.ias.edu/~jnb under the menu item Solar
Neutrinos/History). It is not feasible to present in a twenty
minute talk a balanced account of these three articles with the
appropriate acknowledgments of the important work of so many
people. Therefore, I shall just describe some of the highlights
regarding the standard solar model from a very personal view. I
encourage the listeners who are interested in a more balanced
presentation to look back at the earlier articles which provide
references to critical work done by a large number of researchers.

\section{The development of the ``standard solar model" for
neutrino predictions} \label{sec:development}

I describe the development of the ``standard solar model" for
neutrino predictions in four subsections, covering the period
1962-1988 (Section~\ref{subsec:6268}), 1988-1995
(Section~\ref{subsec:8895}), 1995-1997
(Section~\ref{subsec:9597}), and 1998-2002
(Section~\ref{subsec:9802}).
\subsection{1962-1988}
\label{subsec:6268}

 At the time Ray and I first
began discussing the possibility of a solar neutrino experiment,
in 1962, there were no solar model calculations of solar neutrino
fluxes. Ray, who heard about some of my work on weak interactions
from Willy Fowler, wrote and asked if I could calculate the rate
of the $^7$Be electron capture reaction  in the Sun.

After I did the calculation and submitted the paper to Physical
Review, I woke up to the obvious fact that we needed a detailed
model of the Sun (the temperature, density, and composition
profiles) in order to convert the result to a flux that Ray might
consider measuring. I moved to Willy's laboratory at  CalTech,
where there were experts in stellar modeling who were working on
stellar evolution. We used the codes of Dick Sears and Icko Iben,
and a  bit of nuclear fusion input that I provided, to calculate
the first solar model prediction of solar neutrinos in
$1962-1963$.

The result was extremely disappointing to Ray and to me, since the
event rate from neutrino capture by chlorine that I calculated
from our first flux evaluation was too small by an order of
magnitude to be measured in any chlorine detector that Ray thought
would be feasible. The situation was reversed in late 1963, when I
realized that the capture rate for $^8$B neutrinos on chlorine
would be increased by almost a factor of $20$ over my earlier
calculations because of transitions to the excited states of
argon, most importantly the super-allowed transition from the
ground state of $^{37}$Cl to the isotopic analogue state at about
$5$ MeV excitation energy in $^{37}$Ar.  This increase in the
predicted rate made the experiment appear feasible and Ray and I
wrote a joint paper for Physical Review Letters proposing a
practical chlorine experiment, a paper that was separated into two
shorter papers to meet the space requirements.

During the period $1962-1968$, the input data to the solar models
were refined in a number of important ways as the result of the
hard work of many people. The most significant changes were in the
measured laboratory rate for the $^3$He-$^3$He reaction (changed
by a factor of 3.9), in the theoretically calculated rate for the
$p-p$ reaction (changed by $7$\%), and the observed value of the
heavy element to hydrogen ratio, $Z/X$ (decreased by a factor of
$2.5$). Unfortunately, each of the individual corrections were in
a direction that decreased the predicted flux.

Ray's first measurement was reported in PRL in 1968. Our
accompanying best-estimate solar model prediction (made together
with N. A. Bahcall and G. Shaviv) was about a factor of 2.5 times
larger than Ray's upper limit. But the uncertainties in the model
predictions were, in 1968, sufficiently large that I personally
did not feel confident in concluding that the  disagreement
between prediction and measurement meant that something
fundamental was really wrong.

As it turned out, the values of the stellar interior parameters
used in $1968$ are in reasonably good agreement with the values
used today. However, the uncertainties are much better known now,
after more than three decades of intense and precise studies and
refinements by many different groups working all over the world.

The laboratory  measurement of the $^7$Be($p,\gamma$)$^8$B cross
section was a principal source of uncertainty in the $1962$
prediction, remained a principal uncertainty in $1968$, and is
still today one of the two largest uncertainties in the solar
neutrino predictions. Moreover, the best-estimate measured value
for the cross section has decreased  significantly since 1968 (see
Figure~\ref{fig:be7pcrosssection}).

As we shall see in the subsequent discussion, the only
fundamentally new element that has been introduced in the
theoretical calculations since $1968$ is the effect of element
diffusion in the sun (see 1988-1997 below).

During the period $1968-1988$, very few people worked on topics
related to solar neutrinos. There was only one solar neutrino
detector, Ray's chlorine experiment. His measurement was lower
than our prediction. I concentrated during these two long decades
on refining the predictions and, most importantly, making the
estimates of the uncertainties more formal and more robust.

We calculated the uncertainties by computing the partial
derivatives of each of the fluxes with respect to each of the
significant input parameters. In 1988, Roger Ulrich and I also did
a Monte Carlo study of the uncertainties, which made use of the
fluxes calculated from $1000$ standard solar models. For each of
the $1000$ models, the value of each input parameter was drawn
from a probability distribution that had the same mean and
variance as was assigned to that parameter.  The Monte Carlo
results confirmed the conclusions reached using the partial
derivatives. The uncertainty estimates made during this period are
the basis for the uncertainties assigned in the current neutrino
flux predictions and influence inferences regarding neutrino
parameters (like $\Delta m^2$, $\tan^2 \theta$) that are derived
from analyses that make use of the solar model predictions.

\subsection{1988-1995}
\label{subsec:8895}

In the period $1990-1994$, F. Rogers and J. Iglesias of the
Livermore National Laboratory published their detailed and
improved calculations of stellar radiative opacities and equation
of state. Now almost universally used by stellar modelers, this
fundamental work resolved a number of long standing discrepancies
between observations and predictions of stellar models.

 In the same $1988$ RMP paper in which we presented the Monte Carlo study
of the uncertainties, Roger Ulrich and I also made comparisons
between the predictions of our standard solar model--constructed
to predict solar neutrinos--and the then existing
helioseismological data on $p$-mode oscillations. The agreement
was reasonably impressive: the model predictions and the measured
frequency splittings  agreed to about $0.5$\%. But, we suspected
that there was something missing in the solar models.

During the period $1990-1995$, my colleagues and I made
successively better approximations at including element diffusion
in the solar model calculations. First, we derived an approximate
analytic description which was included in the solar models (after
some significant coding struggles) and later we made use of a
precise computer subroutine that calculated the diffusion
numerically. This work was done with S. Basu, A. Loeb, M.
Pinsonneault, and A. Thoule.

\subsection{1995-1997}
 \label{subsec:9597}
In $1995$, Steve Tomczyk and his colleagues presented the first
observations of the solar $p$-mode oscillations that included
modes that sampled well both the intermediate solar interior and
the deep interior. These observations determined precise
observational values for the sound speed over essentially the
entire solar interior.

We were in a wonderful position to make use of these precise sound
speeds. In $1995$,  Marc Pinsonneault and I had just published a
systematic study of improved solar models that incorporated the
new opacity and equation of state calculations from the Livermore
group and, most importantly, we had succeeded in including helium
and heavy element diffusion in our standard solar model.

\begin{figure}[!t]
\centerline{\psfig{figure=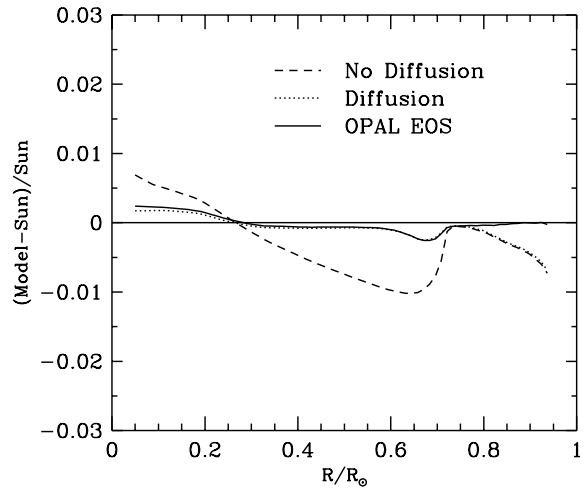,width=3in}} \vglue-.3truein \caption[]{Comparison of measured
and calculated sound speeds. This figure from astro-ph/9610250, PRL {\bf 178},171 1997 compares the sound
speeds calculated with the standard solar model, BP95, with the helioseismologically determined sound
speeds. The dashed curve represents the results from a solar model that does not include element
diffusion. Much better agreement is obtained when element diffusion is included, as indicated by the
dotted curve. The solid line represents a model in which both element diffusion and the refined OPAL
equation of state are included. \label{fig:soundspeeds}}
\end{figure}

Together with Sarbani Basu and Joergen Christensen-Dalsgaard, we
showed that the helioseismologically measured sound speeds were in
excellent agreement throughout the Sun with the values calculated
from our previously constructed standard solar model. As shown in
Figure~\ref{fig:soundspeeds}, the agreement averaged better than
$0.1$\% r.m.s. in the solar interior. We made a simple scaling
argument between accuracy in predicting sound speeds and accuracy
in predicting neutrino fluxes. The concluding sentence in the
Abstract of our PRL paper was:

``Standard solar models predict the structure of the Sun more
accurately than is required for applications involving solar
neutrinos."

This result was published in the January 1997 issue of PRL, but I
had earlier presented at Neutrino '96 in Helsinki (June 1996) the
same conclusion based upon somewhat less precise
helioseismological data. Since many of you were present also at
Helsinki, you may be interested in the precise form of the
statement made in the printed proceedings:

``Helioseismology, as summarized in Figure 2 [a comparison of
measured and calculated sound speeds], has effectively shown that
the solar neutrino problems cannot be ascribed to errors in the
temperature profile of the Sun."

So, from the astronomical perspective, we have known for six years
that new physics was required to resolve the discrepancy between
the standard predictions of the solar model and electroweak
theory. Even prior to the existence of this helioseismological
evidence, it had become clear that one could not fit the data for
all the solar neutrino experiments by simply rescaling standard
predictions of neutrino fluxes.

Why did it take so long? In Section~\ref{sec:why}, I will try to
answer the question: Why were some physicists unconvinced by the
astronomical evidence that solar neutrino oscillations occurred?

\subsection{1998-2002}
\label{subsec:9802}

You have heard beautiful talks at this conference by A. Hallin and
M. Smy on the awesome achievements of the SNO and Super-Kamiokande
collaborations. These experiments have confirmed directly the
calculated solar model flux of $^8$B neutrinos, provided there is
not a large component of sterile neutrinos in the incident flux.

In units of $10^6{\rm ~ cm^{-2}s^{-1}}$, the standard solar model
prediction for the flux, $\phi$, of rare $^8$B neutrinos is

\begin{equation}
\phi({\rm BP00}) ~=~ 5.05^{+1.0}_{-0.8}. \label{eq:bp00prediction}
\end{equation}
In June 2001, the SNO collaboration announced that the combined
result from their initial CC measurement and the Super-Kamiokande
$\nu-e$ scattering measurement implied a flux of $^8$B active
neutrinos equal to

\begin{equation}
\phi({\rm SNO~CC \,+\, SK}) ~=~ 5.44 \pm 0.99. \label{eq:SKplusCC}
\end{equation}
The agreement between the best-estimate calculated value given in
Eq.~(\ref{eq:bp00prediction}) and the best-estimated measured
value given in Eq.~(\ref{eq:SKplusCC}) is  $0.3\sigma$.

The recent SNO NC measurement implies an even closer agreement
between the best-estimates. Assuming an undistorted $^8$B neutrino
spectrum (a very good approximation), the SNO collaboration finds

\begin{equation}
\phi({\rm NC}) ~=~ 5.09 \pm 0.64. \label{eq:SNOnc}
\end{equation}

The agreement between the best-estimates given in
Eq.~(\ref{eq:bp00prediction}) and Eq.~(\ref{eq:SNOnc}) is
embarrassingly small, $0.03\sigma$, but obviously accidental. The
quoted errors, theoretical and experimental, are real and
relatively large.

We shall now discuss uncertainties in the predictions of the solar
neutrino fluxes.

\section{Uncertainties in the solar model predictions}
\label{sec:uncertainties} I will  begin the discussion of
uncertainties with a brief introduction in
Section~\ref{subsec:skepticism} that emphasizes the importance of
robust and well-defined estimates of the errors. Then I will
describe in Section~\ref{subsec:currentuncertainties} the most
important sources of uncertainties in the contemporary
predictions.
\subsection{Skepticism}
\label{subsec:skepticism}

From the very beginning of solar neutrino research, the
uncertainties in the solar model predictions have been a central
issue. If, as many physicists initially believed, the astronomical
predictions were not quantitatively reliable, then there was no
real ``solar neutrino problem."

Even Bruno Pontecorvo, in his prophetic paper ``Neutrino
Experiments and the Problem of Conservation of Lepton Charge",
Soviet Physics JETP, 26, 984 (1968), expressed the view that the
uncertainties in the solar model calculations were so large as to
prevent a useful comparison with solar neutrino experiments, Here
is what Bruno said:

``From the point of view of detection possibilities, an ideal
object is the sun...Unfortunately, the weight of the various
thermonuclear reactions in the sun, and the central temperature of
the sun, are insufficiently well known in order to allow a useful
comparison of expected and observed solar neutrinos, from the
point of view of this article."

This comment by Bruno Pontecorvo is indicative of the skepticism
about solar model predictions that existed among many physicists.
In an effort to assess the validity of this skepticism, I spent
much of the period 1968-2002 investigating the robustness of the
solar model predictions of neutrino fluxes. Even after the
experimental confirmation of the predicted active $^8$B neutrino
flux, the uncertainties in the other solar neutrino fluxes remain
an important ingredient in the determination of neutrino
parameters ($\Delta m^2$, $\tan^2 \theta$) from global analyses of
solar neutrino experiments.

I want to summarize for you now the current best estimates for the
uncertainties in the solar neutrino predictions.

\subsection{Currently estimated uncertainties in predicted neutrino fluxes}
\label{subsec:currentuncertainties} I will first present in
Section~\ref{subsubsec:totalpartial} the current values for the
total and the partial uncertainties in the flux predictions. Then
I will describe in Section~\ref{subsubsec:saga} and
Section~\ref{subsubsec:nucrosssection}, respectively, the very
different histories for the determination of the cross sections
for the  $^7$Be(p,$\gamma$)$^8$B reaction and the $^{37}$Cl($\nu$,
$e^-$)$^{37}$Ar.

\subsubsection{Total and fractional uncertainties}
\label{subsubsec:totalpartial}

\begin{figure}
\centerline{\psfig{figure=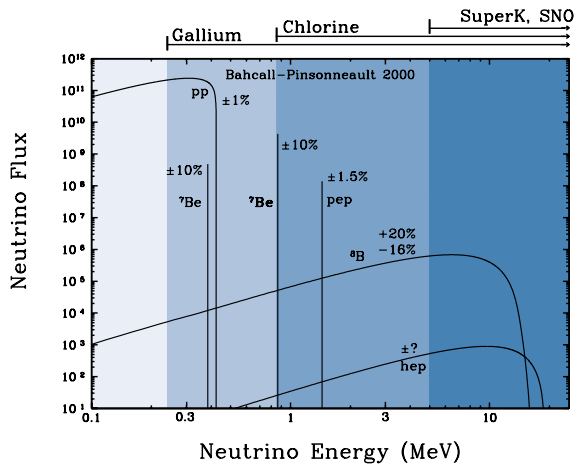,width=3.5in,angle=270}}
\vglue-.3truein \caption[]{Solar neutrino spectrum with currently
estimated uncertainties. \label{fig:snspectrum}}
\end{figure}

 Figure~\ref{fig:snspectrum} shows
the calculated values for the principal (p-p)  solar neutrino
fluxes and their estimated uncertainties. The p-p and pep neutrino
fluxes are predicted with a calculated uncertainty of only $\pm
1$\% and $\pm 1.5$\%, respectively. The $^7$Be neutrino flux is
predicted  with an uncertainty of $\pm 10$\% and the important
$^8$B neutrino flux, which is measured by Super-Kamiokande and
SNO,  is predicted with an error of about $20$\%. The fluxes from
CNO reactions, especially $^{13}$N and $^{15}$O neutrino fluxes,
are predicted with less precision than the fluxes from the p-p
reactions. I have not shown the CNO fluxes in
Figure~\ref{fig:snspectrum} since these fluxes are not expected to
play a discernible role in any of the planned or in progress solar
neutrino experiments.

\begin{table}[!htb]
\centering \caption[]{Fractional uncertainties in the Predicted
$^8$B and $^7$Be  Solar Neutrino Fluxes (BP00). The table presents
the fractional uncertainties in the calculated $^8$B and $^7$Be
neutrino fluxes, due to the different factors listed in the column
labeled Source. The first four rows refer to the low energy cross
section factors for different fusion reactions. The last four rows
refer to the heavy element to hydrogen ratio, $Z/X$ (Composition),
the radiative opacity, a multiplicative constant in the expression
for the diffusion rate of heavy elements and helium, and the total
solar optical luminosity. \label{tab:uncertainties}}
\begin{tabular}{ccc}
\hline\hline
\noalign{\smallskip}
Source& $^8$B&$^7$Be\\
\noalign{\smallskip}
\hline
\noalign{\smallskip}
p-p&0.04&0.02\\
$^3$He+$^3$He&0.02&0.02\\
$^3$He+$^4$He&0.08&0.08\\
p + $^7$Be&$^{+0.14}_{-0.07}$&0.00\\
Composition&0.08&0.03\\
Opacity&0.05&0.03\\
Diffusion&0.04&0.02\\
Luminosity&0.03&0.01\\
\noalign{\smallskip}
\hline\hline
\end{tabular}
\end{table}

Table~\ref{tab:uncertainties} shows how much each of the principal
sources of uncertainty contribute to the total present-day
uncertainty in the calculation of the $^8$B and $^7$Be solar
neutrino fluxes. The largest uncertainty in the prediction of the
$^8$B neutrino flux is caused by the estimated error in the
laboratory measurement of the low energy cross section for the
$^7$Be(p,$\gamma$)$^8$B reaction (This statement was also true in
1962, 1964, 1968, ....). The largest uncertainty in the prediction
of the $^7$Be neutrino flux is due to the quoted error in the
measurement of the low energy rate for the $^3$He + $^4$He
reaction.  In addition, there are a number of other sources of
uncertainty, all of which contribute more or less comparably to
the total uncertainty in the prediction of the $^7$Be and the
$^8$B neutrino fluxes.

\begin{figure}[!t]
\centerline{\psfig{figure=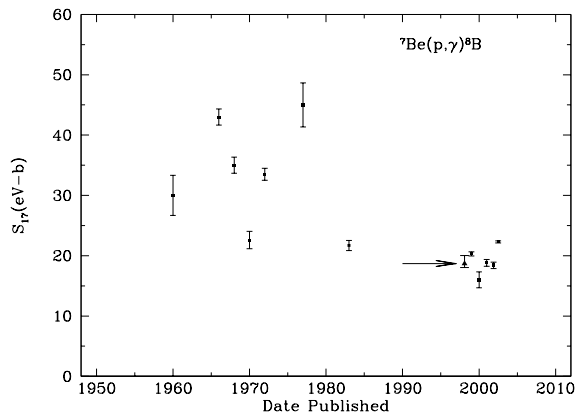,width=3.5in,angle=270}}
\vglue-.3truein \caption[]{$^7$Be(p,$\gamma$)$^8$B. The figure
shows the measured values as a function of date of publication for
the low-energy cross section factor for the
$^7$Be(p,$\gamma$)$^8$B reaction. The arrow points to the
currently standard value, recommended by Adelberger et al., that
is used in the BP00 calculations. \label{fig:be7pcrosssection}}
\end{figure}

\subsubsection{The saga of the $^7$Be(p,$\gamma$)$^8$B cross section}
\label{subsubsec:saga}

 Figure~\ref{fig:be7pcrosssection} shows, as a
function of the date of publication,  the measured values for the
low energy cross section of the crucial reaction
$^7$Be(p,$\gamma$)$^8$B. I have only shown here the direct
measurements of this reaction; there are also indirect
measurements that yield similar results.

 The
encouraging aspect of Figure~\ref{fig:be7pcrosssection} is that
the huge uncertainty that existed between 1960 and 1980, of order
a factor of two, has been much reduced in the following two
decades. In the BP00 calculations, we adopted as the best-estimate
the Adelberger et al. [RMP, 70, 1265 (1998), astro-ph/9805121]
consensus value for the cross section factor of the
$^7$Be(p,$\gamma$)$^8$B reaction, $S_{17}(0) = 19 (1 +
^{+0.14}_{-0.07})$ eV-b (the $1\sigma$ error given here is
one-third the  Adelberger et al. $3\sigma$ estimate). This value
is indicated in the figure by arrow next to ``Standard."

Several refined experiments are in progress or are planned to
measure more accurately the low energy cross section factor for
the $^7$Be(p,$\gamma$)$^8$B reaction or the
p($^7$Be,$\gamma$)$^8$B reaction. Also, there are a number of
related reactions that are being studied in order to give somewhat
more indirect information about the low energy cross section. The
goal of all these experiments is to reduce the combined systematic
and statistical errors to below 5\%, so that $S_{17}(0)$ is no
longer a dominant source of uncertainty in the prediction of the
$^8$B solar neutrino flux (cf. Table~\ref{tab:uncertainties}
above).

To the best of my knowledge, the preliminary data from all of the
existing experiments are consistent with the currently standard
value of $S_{17}(0)$ quoted above. In order to avoid the confusion
that would be created by introducing numbers in the literature
that are changed frequently,  I prefer not to revise the
``standard" estimate of $S_{17}(0)$ (and the $^8$B solar neutrino
flux)  until the in-progress experiments on
$^7$Be(p,$\gamma$)$^8$B and related reactions are completed.

\subsubsection{The $^{37}$Cl($\nu_e$,
$e^-$)$^{37}$Ar cross section} \label{subsubsec:nucrosssection}

\begin{figure}
\centerline{\psfig{figure=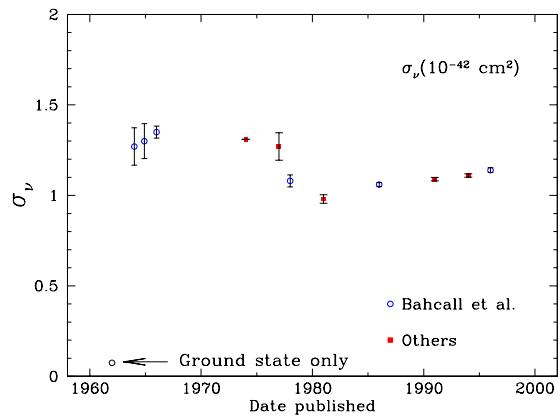,width=3.5in,angle=270}}
\vglue-.4truein \caption[]{$^{37}$Cl($\nu_e$,$e^-$)$^{37}$Ar. The
figure shows, for an undistorted $^8$B solar neutrino spectrum,
the calculated values for the cross section
$^{37}$Cl($\nu_e$,$e^-$)$^{37}$Ar as a function of date of
publication.  \label{fig:nucrosssection}}
\end{figure}

 In the early days of solar neutrino
astronomy, the cross sections for neutrino absorption by chlorine,
$^{37}$Cl($\nu_e$, $e^-$)$^{37}$Ar, were an important source of
uncertainty. For comparison with
Figure~\ref{fig:be7pcrosssection}, I show in
Figure~\ref{fig:nucrosssection} the calculated values of the
absorption cross section for $^8$B neutrinos incident on
$^{37}$Cl. The first calculation I made (in 1962) was too small,
because I did not consider transitions to excited states. The
calculation I made in 1964 was quickly confirmed by measurements
made (by Poskanzer et al.) on the predicted decay: $^{37}$Ca
$\rightarrow$ $^{37}$K + $e^+$ + $\nu_e$, which is the isotopic
analogue of the neutrino capture reaction. A series of subsequent
refined measurements and calculations reduced the estimated error
in the neutrino cross section to where it is no longer one of the
largest sources of uncertainty in the calculation of the predicted
capture rate in the chlorine solar neutrino experiment (although
the uncertainty still plays some role in the global determination
of solar neutrino oscillation parameters).

We shall now discuss the uncertainty in the calculated p-p
neutrino flux.

\section{Is the flux of p-p neutrinos determined by the solar
luminosity?} \label{sec:ppflux}

The predicted p-p neutrino flux is NOT determined by the solar
luminosity independent of details of the solar model. There are
many statements in the literature that make the opposite claim,
namely, that one can calculate the p-p solar neutrino flux without
using a detailed solar model\footnote{One of the most important
recent experimental papers on solar neutrinos states in the
Introduction: ``If we exclude exotic hypotheses, the rate of the
p-p reaction is directly related to the solar luminosity..."}.
These claims are wrong.

Usually, I have managed to keep quiet about this question because
it seemed to be only of academic importance. But, now that p-p
solar neutrino experiments are being seriously developed, it is
important to consider what the p-p flux really can tell us. We
shall see that the p-p neutrino flux cannot be obtained simply
from the observed solar photon luminosity, but instead is
determined by the temperature, density, and composition profiles
of the present-day solar interior.

\subsection{The CNO cycle does not produce p-p neutrinos}
\label{subsec:cno}

The simplest way to see that the p-p neutrino flux is not
determined by the solar luminosity is to recall the results of
Hans Bethe in his epochal paper on main-sequence nuclear fusion
reactions. Hans  concluded, using a crude stellar model, that the
Sun was powered by the now familiar nuclear reactions that make up
the CNO cycle. If the Sun shines by the CNO reactions, then the
p-p neutrino flux is essentially zero. Q.E.D.

Based upon the results of detailed stellar model calculations, we
now believe that stars slightly heavier than the Sun shine
primarily by the CNO reactions, whereas these reactions are
responsible for only about 1.5\% of the solar luminosity.

\subsection{Why the confusion?}
\label{subsec:whyconfusion}

 What is the origin of the confusion?
Why have so many people erroneously claimed that the solar
luminosity determines the p-p neutrino flux?

I think the most important reason is that the solar luminosity
does determine the total solar neutrino flux, just not the p-p
flux\footnote{Even this statement is only correct if one neglects
the difference in neutrino energy loss between different ways of
burning four protons to make an alpha-particle.} . The basic
reaction by which the Sun and other main sequence stars shine is

\begin{equation}
4 p ~\rightarrow~^4{\rm He} ~+~ 2e^+ ~+~2\nu_e. \label{eq:4phe}
\end{equation}
Lepton conservation guarantees that two neutrinos are produced
every time four protons are burned to form an alpha-particle.
However, lepton conservation does not guarantee that two {\it p-p}
neutrinos are produced. Nuclear fusion may produce, for example,
one p-p neutrino and one $^7$Be neutrino.

If a $^3$He ion fuses with an ambient alpha-particle,
$^3$He($\alpha$,$\gamma$)$^7$Be,  before the reaction
$^3$He($^3$He,2p)$^4$He can occur, then one p-p neutrino and one
$^7$Be or $^8$B neutrino will be produced as four protons are
burned. If instead the $^3$He($^3$He,2p)$^4$He reaction occurs
first, then two p-p neutrinos are produced. Even if we could
ignore the possibility of CNO fusion reactions, the solar
luminosity would not determine the p-p neutrino flux.

The solar luminosity does determine the {\it maximum} possible
flux of p-p neutrinos,$\phi_{\rm Max}({\rm p-p})$, via the
relationship:
\begin{eqnarray}
\phi_{\rm Max}({\rm p-p})~&=&~\frac{2 L_\odot}{4\pi(A.U.)^2E_{\rm p-p}}\nonumber\\
~&=&~6.51\times10^{10}\,{\rm cm^{-2}s^{-1}},\label{eq:ppmax}
\end{eqnarray}
where $L_\odot$ is the solar luminosity (in photons), A.U. is the
average Earth-Sun distance, and $E_{\rm p-p}$ is the energy
released (26.197 MeV) to the star when the p-p chain is
terminated by the $^3$He-$^3$He reaction and only p-p neutrinos
are produced. The flux of p-p neutrinos in the standard solar
model happens to be $0.91\phi_{\rm Max}({\rm p-p})$. But, a
detailed solar model is required to determine where the p-p flux
lies between 0 and 1 times $\phi_{\rm Max}({\rm p-p})$.

\subsection{Using p-p and $^7$Be neutrinos to probe details of solar fusion}
\label{subsec:probe}

 Is there any way of probing the solar
interior and determining experimentally which terminating reaction
of the p-p chain, $^3$He-$^3$He or $^3$He-$^4$He, is faster in the
solar interior and by how much? Yes, there is a way. Solar
neutrino experiments can do just that.

 The ratio R of
the rate of $^3$He-$^3$He reactions to the rate of $^3$He-$^4$He
reactions averaged over the Sun can be expressed in terms of the
p-p and $^7$Be neutrino fluxes by the following simple
relation\footnote{More precisely, $\phi(^7{\rm Be})$ should be
replaced by the sum of the $^7$Be and $^8$B neutrino fluxes in the
denominator of Eq.~(\ref{eq:defnR}).}:

\begin{equation}
R ~\equiv~\frac{<^3{\rm He} + ^4{\rm He}>}{<^3{\rm He} + ^3{\rm
He}>} ~=~\frac{2\phi(^7{\rm Be})}{\phi({\rm pp})~-~\phi(^7{\rm
Be})}. \label{eq:defnR}
\end{equation}

The standard solar model predicts $R = 0.174$.  One of the reasons
why it is so important to measure accurately the total p-p and
$^7$Be neutrino flux is in order to test this detailed prediction
of standard solar models.  The value of R reflects the competition
between the two primary ways of terminating the p-p chain and
hence is a critical probe of solar fusion.

\section{Why did it take so long?}
\label{sec:why}

In the introduction to this talk, I said that I would address the
question of why it took so long, about 35 years, to convince many
physicists that solar neutrino research was revealing something
new about neutrinos. I will now do my best to explain why the
process from discovery to consensus required more than three
decades.

In the early years,  after the very rapid progress between 1964 to
1968, there were many, many things that had to be looked at very
carefully to see if there could be something important that had
been left out of the standard solar models. The values of all of
the (large number of) important input parameters were remeasured
or recalculated more accurately, a variety of imaginative
``non-standard solar models" were examined critically, and
possible instabilities in the solar interior were investigated. It
took about 20 years, 1968-1988, for the collective efforts of many
nuclear physicists, atomic physicists, astronomers, and
astrophysicists to provide a thoroughly explored basis for the
standard model calculations that allowed robust estimates of the
uncertainties in the solar model predictions. Even after this long
struggle with details was mostly complete, it was still necessary
to develop codes that could include the refinement of element
diffusion (which took until 1995). And, presumably, there are
still today even further refinements that are appropriate and
necessary to make to obtain a still more accurate description of
the region in which solar fusion takes place.

My impression is that nearly all particle physicists remained
blissfully unaware of, or indifferent to, the decades of efforts
to make the solar neutrino predictions more robust. Why? Why did
many (but not all) particle physicists not take the ``solar
neutrino problem" seriously?

I think that there were three reasons it took so long for particle
physicists to acknowledge that new physics was being revealed in
solar neutrino research. First, the Sun is an unfamiliar
accelerator. Particle physicists, and most other physicists too,
were skeptical of what astronomers and astrophysicists could learn
about an environment that they could neither visit nor manipulate.
These physicists often had only a newspaper-level understanding of
the observational phenomena that stellar models reproduced and the
constraints they met. Second, physicists who heard  my talks, or
heard other talks on solar neutrinos, were most impressed by the
fact that the $^8$B solar neutrino flux depended on the 25th power
of the central temperature, $\phi(^8{\rm B}) \propto T^{25}$. This
dependence seemed too sensitive to allow an accurate prediction
(an objection which was answered experimentally only by the
helioseismological measurements in 1995 and their successful
comparison in 1996-1997 with standard solar model predictions, see
Section~\ref{subsec:9597}). Third, the simplest interpretation of
the discrepancy between observed and predicted solar neutrino
event rates, vacuum neutrino oscillations, suggested large mixing
angles for the neutrinos. It was widely (but not universally)
agreed among particle theorists that mixing angles in the lepton
sector would be small in analogy with the mixing angles in the
quark sector. The most popular view of particle theorists over
most of the history of solar neutrino research has been that since
quarks and leptons are probably in the same multiplets, they
should have mixing angles of comparable size. This objection to
new solar neutrino physics was removed only when Mikheyev and
Smirnov built upon the earlier work of Wolfenstein to describe the
magic of the MSW effect. Ironically, the small mixing angle (SMA)
MSW solution persuaded a significant number of  physicists that
there might be new physics being revealed by solar neutrino
experiments, although today we know that only large mixing angles
solutions are good fits to all the available solar and reactor
neutrino data.

I think that the spirit with which many particle physicists
regarded solar neutrino research is best expressed by a quotation
from the introduction of a 1990 paper written by H. Georgi and M.
Luke [Nucl. Phys. B, 347, 1 (1990)]. They began their article as
follows:

``Most likely, the solar neutrino problem has nothing to do with
particle physics. It is a great triumph that astrophysicists are
able to predict the number of $^8$B neutrinos to within a factor
of 2 or 3..."

Professor Georgi generously allowed me to quote from his paper and
asked only that I emphasize that it was the dependence on the 25th
power of the temperature that maintained his skepticism even after
the invention of the MSW effect.

\section{A personal confession}
\label{sec:confession}

I close this talk with a confession. This is the first time in 40
years of giving talks about solar neutrinos that it seems to me
that the people in the audience are more confident of the solar
neutrino predictions than I am.

More than 99.99\% of the predicted solar neutrino flux is below 5 MeV. We do not yet have any direct
energy measurements of the flux of this dominant component of the Sun's neutrino spectrum. The formal
error on the predicted p-p neutrino flux is only $\pm 1$\%, but I think realistically it would take
another 3 to 10 years of theoretical and experimental research to make sure that we have got everything
(including the uncertainties) as correct as we can at this level of precision. We have to search hard for
uncertainties at a level that is an order of magnitude smaller than the uncertainties that are
significant for interpreting existing experiments. Moreover, it seems to me that there could still be
surprises in the neutrino physics. Simple neutrino scenarios fit well the existing data, which--with the
exception of the chlorine and gallium radiochemical experiments--all detect only solar neutrinos with
energies above 5 MeV. Perhaps these higher energy data have not yet revealed the full richness of the
weak interaction phenomena.

Although we certainly have reason for the celebratory remarks that
have been made at this conference, I often wonder whether Nature
has some beautiful tricks that are still hidden from us.

I am grateful to many colleagues, who are also close friends, with
whom I have collaborated on this subject over the past 40 years.
We have had a lot of fun together. This work was partially
supported by an NSF grant No. PHY0070928.
\end{document}